\documentclass[preprint]{aastex}

\newcommand{\s}{s$^{-1}$}

\shorttitle{Prograde Orbit of WASP-3b}
\shortauthors{Tripathi, Winn, Johnson, et al.}

\begin{document}

%
\def\ltsima{$\; \buildrel < \over \sim \;$}
\def\lsim{\lower.5ex\hbox{\ltsima}}
\def\gtsima{$\; \buildrel > \over \sim \;$}
\def\gsim{\lower.5ex\hbox{\gtsima}}
%

\bibliographystyle{apj}

\title{
A Prograde, Low-Inclination Orbit for the Very Hot Jupiter WASP-3b$^1$
}

\author{
Anjali Tripathi\altaffilmark{2,7},
Joshua N.\ Winn\altaffilmark{2},
John Asher Johnson\altaffilmark{3,8},
Andrew W.\ Howard\altaffilmark{4},\\
Sam Halverson\altaffilmark{4},
Geoffrey W.\ Marcy\altaffilmark{4},
Matthew J.\ Holman\altaffilmark{5},
Katherine R.\ de Kleer\altaffilmark{2},\\
Joshua A.\ Carter\altaffilmark{2},
Gilbert A.\ Esquerdo\altaffilmark{5},
Mark E.\ Everett\altaffilmark{6},
Nicole E. Cabrera\altaffilmark{3,9}
}

\altaffiltext{1}{Some of the data presented herein were obtained at
  the W.~M.\ Keck Observatory, which is operated as a scientific
  partnership among the California Institute of Technology, the
  University of California and the National Aeronautics and Space
  Administration. The Observatory was made possible by the generous
  financial support of the W.~M.\ Keck Foundation.}

\altaffiltext{2}{Department of Physics, and Kavli Institute for
  Astrophysics and Space Research, MIT, Cambridge, MA 02139;
  {\tt tripathi@mit.edu, jwinn@mit.edu}}

\altaffiltext{3}{Institute for Astronomy, University of Hawaii,
  Honolulu, HI 96822}

\altaffiltext{4}{Department of Astronomy, University of California,
  Mail Code 3411, Berkeley, CA 94720}

\altaffiltext{5}{Harvard-Smithsonian Center for Astrophysics, 60
  Garden Street, Cambridge, MA 02138}

\altaffiltext{6}{Planetary Science Institute, 1700 East Fort Lowell
  Road, Suite 106, Tucson, AZ 85719}

\altaffiltext{7}{Current address:\ Institute of Astronomy, University
  of Cambridge, Madingley Road, Cambridge CB3 0HA, U.K.}

\altaffiltext{8}{Current address:\ Department of Astrophysics,
  California Institute of Technology, MC 249-17, Pasadena, CA 91125}

\altaffiltext{9}{Current address:\ School of Physics, Georgia
  Institute of Technology, 837 State St., Atlanta, GA 30332}

\begin{abstract}

  We present new spectroscopic and photometric observations of the
  transiting exoplanetary system WASP-3. Spectra obtained during two
  separate transits exhibit the Rossiter-McLaughlin (RM) effect and
  allow us to estimate the sky-projected angle between the planetary
  orbital axis and the stellar rotation axis, $\lambda=3.3^{+2.5}
  _{-4.4}$ degrees. This alignment between the axes suggests that
  WASP-3b has a low orbital inclination relative to the equatorial
  plane of its parent star. During our first night of spectroscopic
  measurements, we observed an unexpected redshift briefly exceeding
  the expected sum of the orbital and RM velocities by
  140~m~s$^{-1}$. This anomaly could represent the occultation of
  material erupting from the stellar photosphere, although it is more
  likely to be an artifact caused by moonlight scattered into the
  spectrograph.

\end{abstract}


\section{Introduction}

One of the hottest exoplanets yet discovered is WASP-3b, a giant
planet with a mass of $1.76$~$M_{\rm J}$ that orbits an F7-8V star at
a distance of only 0.032~AU, giving the planet an equilibrium
temperature of 1960~K \citep{pollacco}. This close-in, transiting
system is an ideal target for study with the Rossiter-McLaughlin
effect because the host star is both relatively bright ($V=10.5$) and
rapidly rotating ($v\sin i_\star= 13.4 \pm 1.5$~km~s$^{-1}$), and the
transit occurs with an impact parameter near 0.5. Observations of the
Rossiter-McLaughlin effect yield valuable information about the
system, namely the angle ($\lambda$) between the sky projections of
the planetary orbital axis and the stellar rotation axis.  This
parameter is a basic geometric property of the system and a possible
clue about the processes of planet formation and orbital migration.
Detailed descriptions of the Rossiter-McLaughlin effect and its
applications can be found in \citet{ohta}, \citet{gim06},
\citet{gaudi}, and \citet{fabrycky}, while early examples of such
measurements for exoplanetary systems are given by \citet{queloz} and
\citet{winn05}.

In this paper we report the results of new photometric and
spectroscopic observations of WASP-3, which were conducted with the
primary goal of determining $\lambda$ and the secondary goal of
refining estimates of other system parameters. In \S~\ref{sec:obs} we
describe the observations and data reduction, and in
\S~\ref{sec:analysis} we present the model that was used to fit the
data and our results for the system parameters. These results are
discussed in \S~\ref{sec:conc}.

\section{Observations and Data Reduction}\label{sec:obs}

\subsection{Photometric Measurements}

Photometric observations of transits were conducted on UT~2008~May~15,
June~10, June~21, and UT 2009~May~12 and May~25, using Keplercam on the
1.2m telescope at the Fred L.\ Whipple Observatory (FLWO) on Mount
Hopkins, Arizona \citep{holman06}. On each night our observations
spanned the entire transit, except for the first and second nights
when we missed the transit ingress.  For the observations in 2008 we used a
Sloan $i'$ filter, and for those in 2009 we used a Sloan $g'$ filter.
We processed the Keplercam images with standard IRAF procedures for
bias subtraction and flat-field division. Aperture photometry was then
performed on WASP-3 and 6-20 nearby comparison stars, and the WASP-3
signal was divided by a normalized sum of the comparison-star signals.

We also observed the complete transit of UT~2008~July~13 with the
University of Hawaii 2.2m (UH~2.2m) telescope on Mauna Kea, Hawaii. We
used the Orthogonal Parallel Transfer Imaging Camera (OPTIC), which is
equipped with two Lincoln Labs CCID128 orthogonal transfer array (OTA)
detectors \citep{tonry97}. Each OTA detector has 2048$\times$4096
pixels and a scale of 0\farcs135 pixel$^{-1}$.  We took advantage of
the charge-shifting capability of the OTAs to create square point
spread functions (PSF) with side lengths of 39 pixels.  This allowed
us to collect more light before reaching the saturation limit than is
possible in normal imaging mode \citep{howell03, tonry05}. We observed
through a Sloan $z'$ filter, and bias subtraction, flat-fielding, and
aperture photometry were performed with custom IDL procedures
described by \citet{johnson09}.

\begin{figure*}
\epsscale{1}\plotone{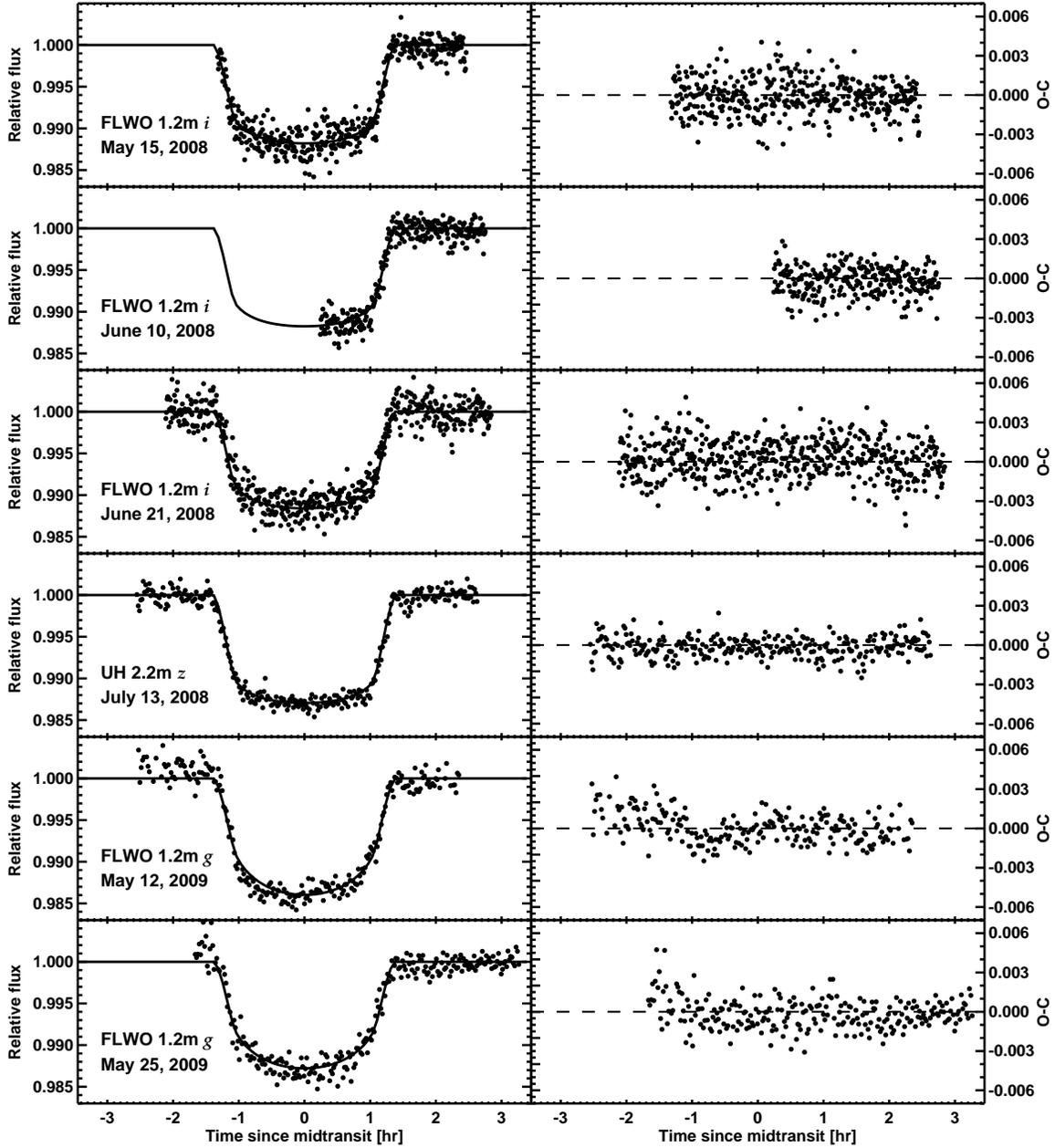}
\caption{Photometry of transits of WASP-3, based on observations with
  the FLWO~1.2m and UH~2.2m telescopes. {\it Left.}---The data and
  best-fitting model. {\it Right.}---Residuals between the data and
  best-fitting model.\label{fig:photfit}}
\end{figure*}

Final light curves are shown in Fig.~\ref{fig:photfit}, after
correcting for differential extinction as described in
\S~\ref{sec:photmodel}.  The photometric data are given in Table
\ref{tab:phot}. 

\subsection{Radial Velocity Measurements}\label{sec:rvmeas}

We measured the apparent radial velocity (RV) of WASP-3 during the
transits of UT~2008~June~19 and 21 with the High Resolution Echelle
Spectrometer (HIRES; Vogt et al.~1994) on the Keck I 10m telescope at
the W.~M.\ Keck Observatory on Mauna Kea. The second of these transits
was observed simultaneously with the FLWO~1.2m telescope as described
in the previous section. We also observed the system with Keck/HIRES
on several other nights in 2008 and 2009, all of which were outside of
transits except for a single measurement obtained during the transit
of UT~2009~June~3.

Our data collection followed the procedure of \citet{johnson08} in
their measurements of HAT-P-1, and the spectrograph was in the same
configuration as used for the California Planet Search \citep{howard,
  marcy}. Calibration of the instrumental response and wavelength
scale was achieved using an I$_2$ gas absorption cell. The slit width
was set by the $0\farcs86$ B5 decker, and the exposure time ranged
from 3 to 5~min, giving a resolution of about 60,000 at 5500~\AA~and a
signal-to-noise ratio of approximately 120~pixel$^{-1}$. Doppler
shifts were derived from the data using the algorithm of
\citet{butler} with subsequent improvements. For a given spectrum,
measurement errors were derived from the weighted standard deviation
of the mean among the solutions for individual 2~\AA~spectral
segments. The measurement error ranged from 6--10~m~s$^{-1}$.

\begin{figure}
\epsscale{1}
\plotone{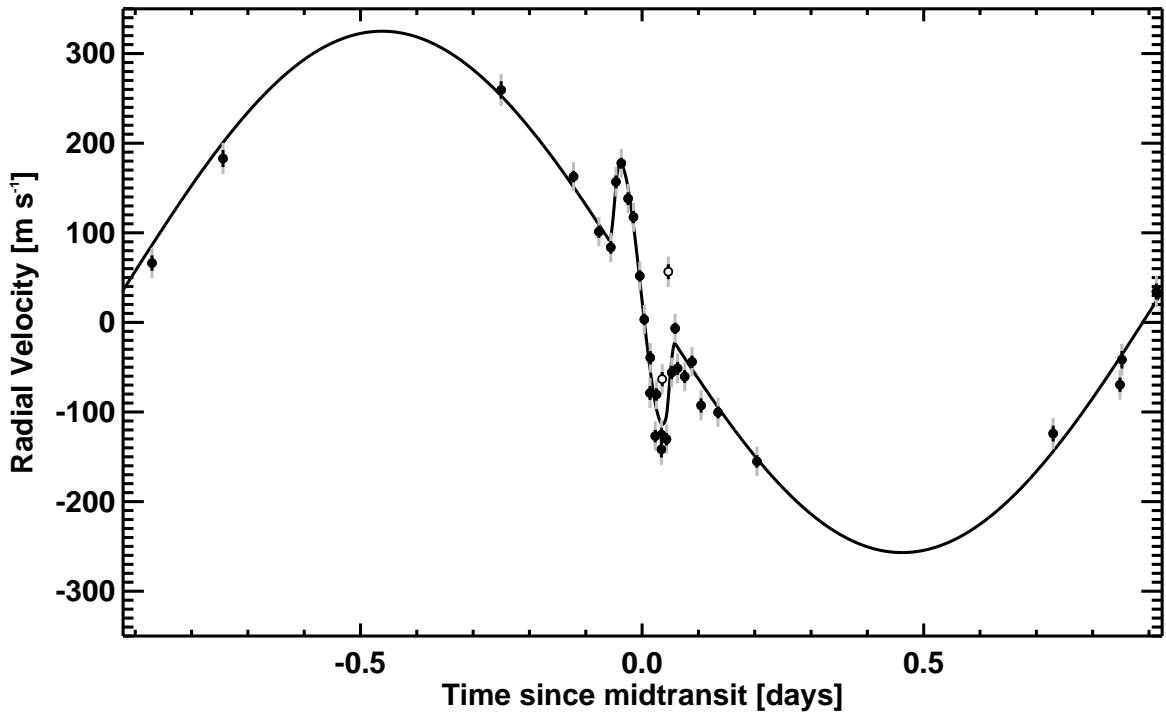}
\caption{
Apparent radial velocity of WASP-3, based on observations with
Keck/HIRES.  The enlarged (gray) error bars include the ``stellar
jitter'' term of 14.8~m~s$^{-1}$ that was added in quadrature with the
measurement errors. The small (black) error bars represent the
measurement error estimated internally from each particular spectrum.
The open symbols represent the ``RV spike.''
  \label{fig:alltime}}
\end{figure}

The RV data are given in Table \ref{tab:keckdata} and plotted in
Fig.~\ref{fig:alltime}. The figures show the expected sinusoidal
variation outside of transits, due to the line-of-sight component of
the Keplerian orbital velocity of the star, as well as the
shorter-timescale anomalous radial velocity due to the
Rossiter-McLaughlin (RM) effect that occurs during the transit phase.

In addition to these two sources of RV variability, an unexpected
redshift was observed near the egress of the UT~2008~June~19
transit. The RV measured at HJD~2,454,637.001983 is redshifted by
140~m~s$^{-1}$ with respect to our best-fitting model (described in
\S~\ref{sec:rvmodel}). This is much larger than both the estimated
measurement error of 8.3~m~s$^{-1}$ and the scatter of
$\approx$15~m~s$^{-1}$ between the other RV data and the best-fitting
model. The previous RV data point, at HJD~2,454,636.991034, is not as
clearly discrepant but it is the second-largest outlier, with a
redshift of 49~m~s$^{-1}$ with respect to the best-fitting model. In
what follows we will refer to these two consecutive RV data points as
the ``RV spike.'' They are identified with open symbols in
Fig.~\ref{fig:alltime}. Casual inspection of the spectra constituting
the spike revealed nothing unusual. In particular, the visual
appearance of the 2-d images, the estimated measurement errors, the
signal-to-noise ratio of the spectra, and the goodness-of-fit
statistic (and other metrics) returned by the Doppler code were all
within normal ranges. Visual examination of the cross-correlation
functions (CCFs) did not reveal anything unusual about the spike
spectra. The time series of bisector spans was also uninformative, as
it shows the same pattern that was already observed in the Doppler
shifts (the Rossiter effect and an additional anomaly during the
spike). Nevertheless, we believe the spike is an artifact of
contamination of the spectra by moonlight, rather than an
astrophysical phenomenon, based on the following logic:
\begin{enumerate}

\item During our observations, the moon was full and 63$^{\circ}$ away
  from WASP-3. Thin cirrus clouds were noted at approximately the time
  of the RV spike, and the mean count rates during the ``spike''
  observations declined by approximately a factor of 2. (The exposure
  times for these spectra were increased to compensate for the reduced
  count rates.) Thus, the spike spectra were obtained during an
  interval of reduced transparency and enhanced sky brightness,
  leading to a possible order-of-magnitude increase in the fractional
  contamination by moonlight.

\item The moonlight absorption lines would have appeared at a velocity
  of approximately $+10$~km~s$^{-1}$ relative to the WASP-3 absorption
  lines (the difference between the barycentric correction of
  4~km~s$^{-1}$ and the WASP-3 systemic velocity of
  $-5.5$~km~s$^{-1}$). This is within the WASP-3 line profiles, which
  are rotationally broadened to $14$~km~s$^{-1}$. Given the low
  contrast ($\approx$10\%) of the WASP-3 lines relative to the
  continuum, a moonlight contamination of only $\sim$1\% of the total
  light would be sufficient to produce a $\sim$100~m~s$^{-1}$ shift in
  the center-of-gravity of the WASP-3 lines. This level of
  contamination would be visually undetectable in the 2-d images.

\item To assess whether 1\% moonlight contamination is reasonable we
  used the moonlight model of \citet{krisciunas}, which
  takes into account the moon phase, star and moon coordinates, and
  observing bandpass.  According to this model, in the {\it absence}
  of clouds the moonlight flux would have been $\approx$0.1\% of the
  flux of WASP-3.  Since the spike spectra were obtained when looking
  through cirrus, it is plausible that the moonlight fraction
  increased by an order of magnitude to 1\%.

\end{enumerate}

Thus, it seems possible and even likely that a small amount of
scattered moonlight was responsible for the RV spike. We cannot be
certain of this conclusion since we could not devise any conclusive
statistical test for moonlight at such low levels, given the
simultaneous presence of the RM effect and the time variability of the
instrumental broadening profile and the CCFs. For our analysis we
omitted the RV spike data from consideration, although we discuss
another possible interpretation in \S~\ref{sec:conc}. The
investigation of the RV spike was the purpose of the coordinated
spectroscopic and photometric observations of the transit of
UT~2008~June~21. The RV spike did not recur, and no outliers of
similar magnitude were observed.

\section{Analysis and Results}\label{sec:analysis}

\subsection{Photometric Model} \label{sec:photmodel}

The first step in our analysis was to fit a parametric model to the
photometric data. We simplified the analysis by assuming that the
planet's orbital eccentricity is zero, which is consistent with
previous results \citep{pollacco,madhusudhan} as well as our own RV
observations and the expectation that such a close-in orbit has been
tidally circularized. The orbit was parameterized by the period ($P$),
the midtransit time ($T_c$), the impact parameter ($b$) in units of
the stellar radius, the planet-to-star radius ratio ($r \equiv
R_p/R_\star$), and the approximate transit duration ($T$), defined in
\citet{carter08} as
\begin{equation}
T=\frac{P}{\pi b} \sqrt{1-b^2} \cos{i},
\label{eqn:T}
\end{equation}
where $i$ is the inclination angle.

The loss of light during the transit was calculated using the analytic
formulae of \citet{mandel} for a quadratic limb-darkening (LD) law
with coefficients $u_1$ and $u_2$. To speed the convergence of our
fitting algorithms, we used the nearly-uncorrelated linear combinations of
the LD coefficients,
\begin{eqnarray}
u_1' & = & u_1\cos\phi + u_2\sin\phi,\\
u_2' & = & u_2\cos\phi - u_1\sin\phi,
\end{eqnarray}
with $\phi=39^{\circ}$ \citep{pal08}. As a compromise between allowing
complete freedom in the LD law and placing complete trust in a
theoretical LD law, we allowed $u_1'$ to be a free parameter and fixed
$u_2'$ at the value predicted from the PHOENIX atmosphere models of
\citet{claret04}.  Additionally as in \citet{carter09}, we required $u_1>0$ and
$0<u_1+u_2<1$, to guarantee that the intensity profile of the star
decreased monotonically toward the limb.

In addition, we included two additional free parameters for each night
of observations, to specify the out-of-transit magnitude ($m_{\rm
  oot}$) and a coefficient ($k$) for differential airmass extinction,
\begin{equation}
m_{\rm obs} = m_{\rm oot} + \Delta m + kz,
\end{equation}
where $m_{\rm obs}$ is the observed magnitude, $\Delta m$ is the
calculated magnitude change according to the \citet{mandel}
formulae, and $z$ is the airmass. As with the LD coefficients, it
proved advantageous to use fitting parameters that were appropriate
linear combinations of $m_{\rm oot}$ and $k$.

We fit the data from all 6 transits, allowing each of the 6 time
series to have independent values of $T_c$, $r$, $k$, and $m_{\rm
  oot}$. We required consistency in $b$ and $T$, and in the LD
coefficients for all those data sets observed through the same
bandpass. However, we did {\it not} require consistency in $r$, in
order to check for variations in the transit depth that could be
caused by a time-variable pattern of starspots or other localized
intensity variations across the stellar disk. A time-variable pattern
of such irregularities would lead to variations in the transit depth
from event to event. For example, when there are dark spots located on
the visible hemisphere of the star, the fractional loss of light
during transit is greater than when the star is spotless. These
transit-to-transit variations would be manifested in our model as an
apparent change of the $r$ parameter, even though the actual radii of
the bodies are not changing.  The idea was to seek any evidence for
stellar activity, which would also provide an alternate explanation
for the RV spike.

To determine the best values of the parameters and their
uncertainties, we used a Markov Chain Monte Carlo (MCMC) algorithm,
implemented as described by \citet{holman06} and
\citet{winn07}. Specifically we used a Gibbs sampler and tailored our
jump sizes so that they yielded acceptance rates of approximately 40\%
for each parameter.  The number of links in each of our chains was
approximately $10^5$ per parameter, so that the distributions for each
parameter were well converged.  Individual jumps were executed with
probability $\exp(-\Delta\chi_f^2/2)$, where
\begin{equation}
\chi^2_{f} = \sum_{i=1}^{N_f}\left(\frac{f_i( \textnormal{obs} ) - f_i( \textnormal{calc} )}{\sigma_i}
\right)^2.
\label{eqn:chif}
\end{equation}
In this expression, $N_f$ is the number of data points, $f_i
(\textnormal{obs})$ is the observed flux, $f_i(\textnormal{calc})$ is
the flux that is calculated for a given set of model parameters, and
$\sigma_i$ is a constant for each light curve, determined as
follows. First, a value $\sigma_0$ was determined such that when using
$\sigma_i=\sigma_0$, $\chi^2_f = N_{\rm dof}$ for the best-fitting
model. (The answer was always close to the standard deviation of the
out-of-transit data.)  Next, we multiplied $\sigma_0$ by a factor
$\beta$ to account for time-correlated errors, using the
time-averaging method of \citet{pont06} as implemented by
\citet{winnxmas}. Averaging times of 10--30~min were used to compute
$\beta$, giving results (in chronological order) of 0.92, 1.21, 1.47,
1.30, 2.45, and 1.67. When $\beta$ was found to be smaller than unity,
as in the first case, we used $\beta=1$, reasoning that time
correlations can only increase errors. Finding $\beta<1$ represents
either a statistical fluke or a signal in which the red noise has been
underestimated by analyzing the model residuals rather than the
out-of-transit data (see \citet{carterwinn09} for a discussion).

The MCMC analysis was conducted in several stages, taking advantage of
the fact that the subset of parameters $\{T_c, k, m_{\rm oot}\}$ are
nearly uncorrelated with the other parameters in the model.  First, we
determined the midtransit time ($T_c$) of each event by allowing
$T_c$, $k$, and $m_{\rm oot}$ to be free parameters and fixing all
other parameters (which are uncorrelated with $T_c$) at the values
that minimize $\chi^2_f$. The results for the midtransit times are
given in Table~\ref{tab:tc}, along with other previously reported
midtransit times.

\begin{figure}
\epsscale{1.0}
\plotone{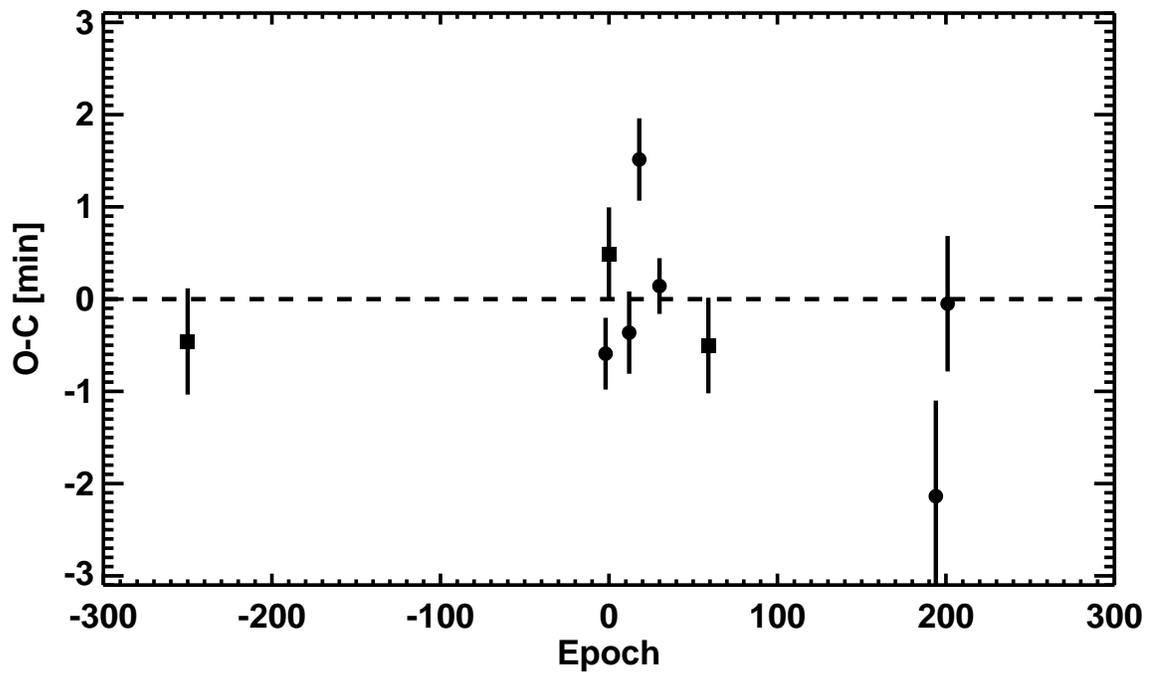}
\caption{ 
Transit timing residuals. A linear function of epoch was fit to the
transit times given in Table~\ref{tab:tc}, and the calculated times
were subtracted from the observed times.  Circles denote midtransit
times from this work; squares denote midtransit times from earlier
studies \citep{pollacco, gibson}.\label{fig:ominusc}}
\end{figure}

Our collection of midtransit times was then used to determine the
transit ephemeris. The results are given in
Table~\ref{tab:photvalues}.  The linear fit that was used to calculate
the ephemeris yielded $\chi^2 = 21$ with 7 degrees of freedom,
indicating an unacceptable fit. The transit timing residuals from this
fit are shown in Fig.~\ref{fig:ominusc}. Given the large value of
$\chi^2$, there are either genuine period variations, or the transit
time uncertainties have been underestimated in some cases. To be
conservative for the purpose of planning future observations, we have
increased the errors of our ephemeris values by a factor of
$\sqrt{21/7}$ above the formal errors of the linear fit.

We determined other system parameters by computing another chain,
using our newly determined ephemeris, holding $T_c$, $k$, and $m_{\rm
  oot}$ fixed at the best-fitting values for each light curve, and
allowing all other parameters to vary. The key results are given in
Table~\ref{tab:photvalues}.

Most notable, as mentioned earlier, is the considerable variation in
the planet-to-star radius ratio ($r$), at the 2--3$\sigma$ level
between the different transits.  The mean value of $r$ across all 6
transits is $0.1059$, from which the individual results differ by as
much as $5\sigma$.  The variation in $r$ can also be examined in terms
of transit depth ($\delta \equiv r^2$).  The mean of the 6 $\delta$
values is 0.01122, and the standard deviation is 0.00080, which is
larger than the statistical uncertainty in any individual depth
measurement.  As an estimate of the transit-to-transit variation in
the transit depth, we sought the value of $\sigma_{\delta}$ such that
\begin{equation}
\sum_{i=1}^{6} \frac{\left(\delta_i-0.01122 \right)^2}
                   {\sigma_i^2 + \sigma_{\delta}^2} = 5,
\end{equation}
with the result $\sigma_\delta = 0.00075$. Thus, the data are
consistent with fractional variations in the transit depth of order
$\sigma_\delta / \delta \approx 7\%$.

Unfortunately we cannot be confident in the reality of transit depth
variations because of the possibility of systematic errors. The light
curves shown in Fig.~\ref{fig:photfit} clearly exhibit red noise,
especially the $g'$-band light curves. Our fitting procedure attempts
to take the red noise into account but it is inevitably imperfect. It
is also suggestive that the $i'$-band light curves gave consistent
results for $\delta$, while the $z'$ and $g'$ light curves gave larger
values for $\delta$; this might be due to systematic errors in the
treatment of LD.

Comparison with previous studies reveals that we found a larger
transit depth than \citet{pollacco} and, similarly, a larger radius
ratio than \citet{gibson}.  Our results for these parameters disagree
with each of the two studies by 3$\sigma$ and 5$\sigma$, respectively.
Of our other photometric parameters, we find agreement with the values
that were reported by \citet{pollacco}, but less so with those
reported by \citet{gibson}. The latter authors reported a more precise
result for $b$ of $0.448\pm 0.014$ that disagrees with our result of
$0.531^{+0.036}_{-0.043}$. They also found a shorter transit duration,
$2.753^{+0.020}_{-0.013}$~hr, compared to our result of $2.813\pm
0.012$~hr.  The reason for these discrepancies is unclear, but we
suspect that at least part of the reason is that the \citet{gibson} uncertainties were underestimated because the LD law was
held fixed.

\subsection{Radial Velocity Model}\label{sec:rvmodel}

We fitted the RV data with a parametric model that takes into account
both the orbital velocity of the star ($V_o$) and the anomalous
velocity due to the RM effect ($\Delta V_r$). As in our photometric
model, we maintained the assumption of a circular orbit.  Our orbital
model was parameterized by the period ($P$), the midtransit time
($T_c$), the velocity semi-amplitude ($K$), and a constant velocity
offset ($\gamma$). The offset parameter was needed because the precise
Doppler velocities were computed with respect to a template spectrum
with an arbitrary velocity zero point.  Furthermore, as we will
describe shortly, we included an additional velocity offset
($\Delta\gamma$) specific to the observations that were made on
UT~2008~June~19.

The RM effect was parameterized by the projected spin-orbit angle
($\lambda$) and the line-of-sight stellar rotational velocity ($v\sin
i_\star$). To calculate $\Delta V_r$ we used the ``RM calibration''
procedure of \citet{winn05}: we simulated spectra exhibiting the RM
effect at different phases of the transit and then measured the
apparent radial velocity of the simulated spectra.\footnote{The
  starting point of the simulations was a spectrum from
  \citet{allende} of Procyon (F5, $T_{\rm eff}=6500$~K) that was
  chosen because it is a slower rotator than WASP-3 but its spectrum
  is similar in other respects. For details on the construction of the
  simulated Rossiter-McLaughlin spectra, refer to Winn et al.~(2005).}
We found the results to be consistent with the formula
\begin{equation}
\Delta V_r = -(\Delta f) V_p
\left[1.51 - 0.44 \left(\frac{V_p}{10.0~{\mathrm{km~s}}^{-1}}\right)^2 \right],
\end{equation}
where $\Delta f$ is the fractional loss of light during the transit
and $V_p$ is the line-of-sight component of the stellar rotation
velocity at the location hidden by the planet.

Since $\Delta V_r$ depends on $\Delta f$, the RV model is dependent on
the photometric parameters.  To determine these parameters, one could
fit the RV and photometric data simultaneously, but we found it faster
and more convenient to fit the RV data alone using prior constraints
on the relevant photometric parameters. Specifically we used $b$, $T$,
and $r$ as parameters in the RV model, the first two of which were
subjected to Gaussian prior constraints from the photometry. As for
$r$, given the discrepancies among the results of fitting individual
transits, we adopted a Gaussian prior with a central value of 0.1049
and an uncertainty of 0.0037, reflecting the median and standard
deviation of the 6 different results.

Thus, our RV fitting statistic was
\begin{eqnarray}
\chi^2_v = \sum_{i=1}^{N_v}\left(\frac{v_i( \textnormal{obs} ) -
                                    v_i( \textnormal{calc} )}{\sigma_i}\right)^2 +
  \left(\frac{b - 0.531}{\sigma_b}\right)^2 +
  \left(\frac{T - 2.412 ~\textnormal{hr}}{\sigma_T}\right)^2 +
  \left(\frac{r - 0.1049}{0.0037}\right)^2,
\end{eqnarray} 
where $N_v$ is the number of RV data points, $v_i (\textnormal{obs})$
is the observed RV, $v_i(\textnormal{calc})$ is the RV that is
calculated for a given set of model parameters, and $\sigma_b$ and
$\sigma_T$ are taken from Table~\ref{tab:photvalues} depending on the
sign of the difference in the numerator. The LD coefficients were
fixed at the values tabulated by \citet{claret04} for the PHOENIX
model in the $g'$ bandpass (where we find the strongest signal from
the $I_2$ absorption lines used for RV data calibration). The period
and midtransit time were fixed at the values determined in
\S~\ref{sec:photmodel}.

The two measurements constituting the RV spike of UT~2008~June~19,
described in \S~\ref{sec:rvmeas}, were significant outliers that did
not conform to our model, regardless of the parameter values that were
chosen. Assuming that they were influenced by a physical process that
was not described by our model---whether moonlight, as we discussed in
\S~2, or an astrophysical phenomenon---we omitted these two data
points from the fit. We also omitted a third RV point immediately
following the spike, even though this data point was not an outlier,
because it was based on observations through cirrus clouds and was
thus suspected of being affected by moonlight. (None of the results we
describe subsequently are altered materially if this third point is
instead included in the fit.) Furthermore, we allowed the data from
UT~2008~June~19 to have an extra velocity offset ($\Delta\gamma$), in
addition to the velocity offset ($\gamma$) applied to all of the RV
data. We incorporated this additional offset to allow for any
systematic offset of the data from that night, again due to moonlight
or astrophysical phenomenon responsible for the RV spike.

Taking $\sigma_i$ to be the measurement uncertainty given in
Table~\ref{tab:keckdata}, the minimum $\chi_v^2$ was 127 with 28
degrees of freedom, an unacceptable fit. We enlarged the RV errors by
adding 14.8~m~s$^{-1}$ in quadrature with the measurement errors in
order to achieve a reduced $\chi_v^2$ of unity.  The excess scatter is
referred to as ``stellar jitter'' and is attributed to intrinsic
motions of the stellar photosphere, unmodeled orbital motions (due to
additional planets or companion stars) and unknown systematic errors.
This procedure of adding in stellar jitter is common in fitting
high-precision RV data, and the value of 14.8~m~s$^{-1}$~is in line
with previous observations of F stars rotating as rapidly as WASP-3
\citep{saar}.

\begin{figure}
\epsscale{0.8}
\plotone{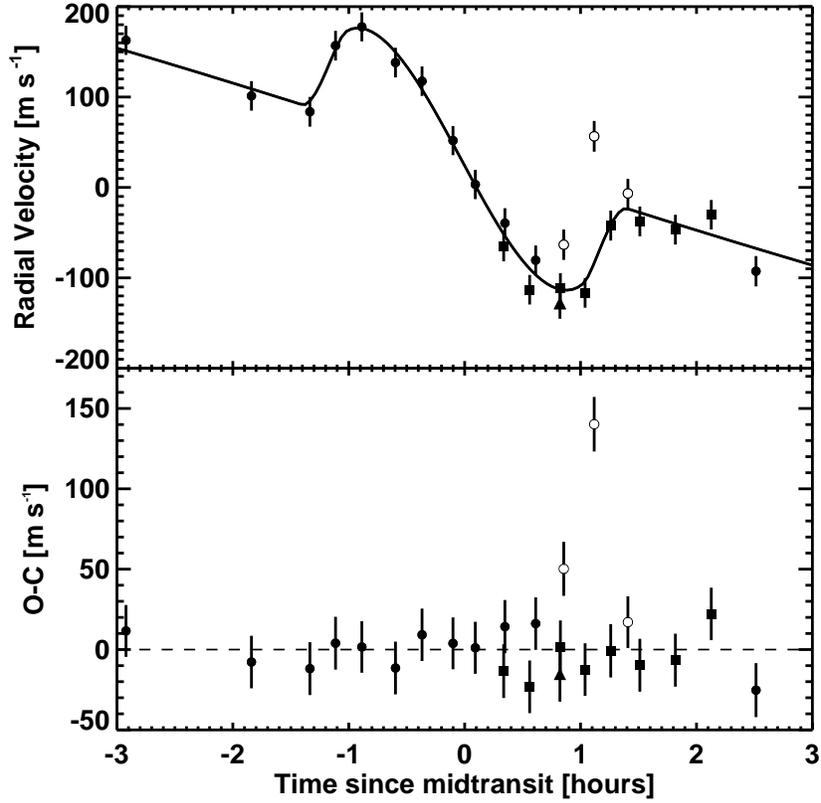}
\caption{
The spectroscopic transit of WASP-3.  {\it Top.}---The RV data from UT
2008~June~19 (circles), 2008~June~21 (squares), and 2009~June~3
(triangle). The open symbols show the data which were omitted from the
fit: the 2 points of the RV spike, and the next data point, which is
also anomalously redshifted although by a much smaller degree. The
constant velocity offsets given in Table~\ref{tab:rvparams} have been
applied and the error bars shown here include stellar jitter.  {\it
  Bottom.}---Residuals between the data and the best-fitting
model.\label{fig:bestfit}}
\end{figure}

Fig.~\ref{fig:bestfit} shows the RV data, with the omitted points
highlighted, the error bars enlarged to encompass the stellar jitter,
and the velocity offsets applied. Results from the fit are given in
Table~\ref{tab:rvparams}.  We find the parameters describing the RM
effect to be $v\sin i_\star=14.1 ^{+1.5} _{-1.3}$ km~\s~and
$\lambda=3.3 ^{+2.5} _{-4.4}$ degrees. Our result for $v\sin i_\star$
is in agreement with the value of $13.4 \pm 1.5$~km~\s~that was
previously reported by \citet{pollacco}, based on an analysis of the
spectral line profiles rather than on the RM effect. The consistency
between these results lends confidence to both analyses and our ``RM
calibration'' procedure.  Our result for $\lambda$ is new information
about this planetary system, indicating good alignment between the
stellar rotation axis and the projected orbit normal. Our result for
the velocity semi-amplitude $K$ is $290.5_{-9.2}^{+9.8}$~m~\s, which
is about 3$\sigma$ larger than the value of
$251.2_{-10.8}^{+7.9}$~m~\s reported by \citep{pollacco}, but which is
in agreement with the more recent determination of $274\pm
11$~m~s$^{-1}$ by \citet{2009arXiv0912.3643S}.

\section{Discussion}\label{sec:conc}

We have observed the Rossiter-McLaughlin effect during transits of
WASP-3b and found that the projected spin-orbit angle is $\lambda=3.3
^{+2.5} _{-4.4}$ degrees, suggesting that the stellar rotation axis
and the orbital axis are closely aligned.\footnote{While the first
  draft of this manuscript was under review, we learned of a similar
  study of WASP-3 by \citet{2009arXiv0912.3643S}.  Those authors also
  measured the RM effect, and found $\lambda=15_{-9}^{+10}$~deg, also
  suggesting a relatively close spin-orbit alignment.} This alignment
makes WASP-3 similar to the majority of other exoplanets whose
alignment has been studied \citep{fabrycky} and unlike the misaligned
systems XO-3 \citep{hebrard, winn-rm-xo3}, HD~80606 \citep{pont09,
  winn-rm-hd80606}, WASP-14 \citep{johnson-wasp14}, HAT-P-7 \citep{narita, winn-rm-hatp7}, Corot-1 \citep{pont09}, and WASP-17 \citep{anderson}. It would seem that for WASP-3, the inward migration process
did not disrupt the initial coplanarity of the system, or tidal
effects have coplanarized the system (although the theoretically
calculated rate of coplanarization is too slow to be relevant,
\citet{barker}).

We also used new RV measurements and transit light curves to determine
the system parameters, and found agreement with previously published
values, with the exception of the RV semi-amplitude $K$ reported by
\citet{pollacco}. In addition, we observed an intriguing RV
anomaly during one of the transits. Our investigation of the spectra
led us to the conclusion that this ``RV spike'' was caused by a small
amount of moonlight that was admitted to the slit along with the light
from WASP-3.  However, we cannot be completely certain in this
interpretation, and in any case it is interesting to ask whether there
are any astrophysical phenomena that could produce such an effect.

One speculative idea is that the spike represents the planet's passage
in front of a starspot.  When a spot is hidden by a planet, the
apparent RV of the star will jump to a larger or smaller value,
because the RM-like effect of the spot temporarily vanishes.  When the
spot is no longer hidden, the star's apparent RV returns to its
expected transiting value.  The resulting ``spike'' in RV would be
analogous to the photometric ``bumps'' or ``rebrightenings'' that have
been seen in transit light curves and attributed to starspot
crossings, as in \citet{rabus} and \citet{dittmann}, amongst others.

The time between the RV spike's initial point and its maximum
($\approx 20$~mins) is consistent with the ingress or egress duration,
as would be required of the spot-occultation hypothesis.  However, to
match the $+140$~m~s$^{-1}$ amplitude of the RV spike, the starspot
would need to present a {\it velocity}\, contrast with the
photosphere, in addition to a possible intensity contrast. This can be
understood as follows: under the starspot hypothesis the amplitude of
the RV spike would have an amplitude
\begin{equation}
\Delta v \approx v_\star
\left(\frac{R_s}{R_\star}\right)^2\left(1 - \frac{I_sv_s}{I_\star v_\star}\right),
\label{eq:rv-noise}
\end{equation}
assuming there is a single spot with radius $R_s$, intensity $I_s$,
and radial velocity $v_s$ that may differ from the surrounding
photosphere. In the absence of the spot, the material at the spot's
location would have intensity $I_\star$ and radial velocity
$v_\star$. Because the spike occurred near the transit egress, the
spot would have been near the receding limb of the star, where
$v_\star = +v\sin i_\star\sqrt{1-b^2} \approx
+12$~km~s$^{-1}$. Assuming $R_s\approx R_p$ and setting $\Delta v$ of
Eqn.~(\ref{eq:rv-noise}) equal to $140$~m~s$^{-1}$, we find
\begin{equation}
\frac{I_s v_s}{I_\star v_\star} \approx -0.06.
\end{equation}
Since intensity cannot be negative, the above relation implies that
the spot's effective RV must be oppositely directed from the local
photospheric RV.  The planet must cover a blueshifted spot in order to
produce a redshift as large as 140~m~s$^{-1}$. Specifically, the
spot's peculiar RV, defined as $v_s - v_\star$, is approximately
$-12.7$~km~s$^{-1}$~$(I_\star/I_s)$.  Unless the spot is very bright
($I_s \gg I_\star$) the peculiar velocity just estimated would exceed
the sound speed of the photosphere, which we estimate to be
7~km~s$^{-1}$. The hidden-spot hypothesis for the RV spike thus
requires not an ordinary spot, but rather an erupting and possibly
even supersonic spot.  What we are calling a ``spot'' would actually
be more akin to an active site undergoing a stellar flare.

Thus, the RV anomaly could be rationalized as a consequence of vigorous
stellar activity, a hypothesis which would also produce transit depth
variations (see \S~3.1). A serious objection to this hypothesis,
however, is that by the usual measures WASP-3 is a quiescent
main-sequence F7-8 dwarf star for which stellar activity is not
expected.  \citet{pollacco} presented no evidence for
photometric variations, chromospheric emission lines, or any other
indication of stellar activity. Likewise, there is no detectable Ca II
H~\&~K emission in any of our own Keck/HIRES spectra, including those
obtained during the RV spike.  The measured Mt.~Wilson $S$ values had
an RMS of only 0.0013, with no noticeable offset even between the
measurements spanning different seasons.  Considering $R'_{HK}$, the
ratio of chromospheric emission in Ca II H \& K cores to the total
stellar emission, we find that WASP-3's mean $\log{R'_{\rm HK}}=-4.9$,
giving it an approximate age of 4 Gyr \citep{noyes}.  By comparison, a
very active star such as HD189733 has $\log{R'_{\rm HK}}=-4.5$, while
a very quiet star such as HD9407 has $\log{R'_{\rm HK}}=-4.98$,
\citep{wright04}. Finally, one might expect an unusually active star
to be an unusually rapid rotator, but the value of $v\sin
i_\star=14.1$~km~s$^{-1}$ is typical of an F7V star. The activity
hypothesis would therefore imply a high level of activity in a star
that one would otherwise expect to be quiescent.

Although the activity hypothesis is not tenable for WASP-3, we 
predict that similar observations of other systems with more active
stars will eventually reveal ``RV spikes'' when the planet occults
active regions.  Simultaneous photometry and spectroscopy of these
events will allow investigations of the intensity and velocity
contrast of the active regions, relative to the surrounding
photosphere.  Active stars are often excluded from Doppler planet
surveys, precisely because the excess RV noise caused by stellar
activity is a hindrance to planet detection.  However,
magnitude-limited photometric transit surveys have no such selection,
and in this regard it is unsurprising that planets around very active
stars are being routinely discovered by these surveys, such as CoRoT-2
\citep{lanza} and CoRoT-7 \citep{leger}.  The Doppler follow-up for
these systems will undoubtedly be more difficult than for quiescent
stars.  But, perhaps these difficulties will be compensated to some
degree by the prospect that active stars with eclipsing planets will
eventually enable a deeper understanding of starspots, stellar
activity, and stellar flares.

\acknowledgments We thank N.\ Madhusudhan, Michael L.\ Stevens, and
Robert Noyes for helpful discussions. We are grateful to the anonymous
referee for helping to clarify the interpretation of the RV spike. The
authors wish to recognize and acknowledge the very significant
cultural role and reverence that the summit of Mauna Kea has always
had within the indigenous Hawaiian community. We are most fortunate to
have the opportunity to conduct observations from this mountain. A.T.\
and K.dK.\ thank the MIT UROP Endowment Fund and Office for financial
support.  J.N.W.\ is grateful to the NASA Origins program for support
through grants NNX09AD36G and NNX09AB33G, and to the MIT Class of 1942
for a Career Development Professorship.  J.A.J.\ is an NSF Astronomy
and Astrophysics Postdoctoral Fellow with support from the NSF grant
AST-0702821.

Facilities: \facility{FLWO:1.2m}, \facility{UH:2.2m},
\facility{Keck:I}.

\clearpage
\begin{deluxetable}{cccc}

\tablewidth{0pt}
\tablecaption{Relative Photometry of WASP-3\label{tab:phot}}
\tablehead{\colhead{Heliocentric Julian Date} & \colhead{Relative Flux} & \colhead{Uncertainty} &\colhead{Observatory}  } 
\startdata
2454601.810140 & 0.99857 & 0.00134 & 1 \\
2454601.810534 & 0.99712 & 0.00134 & 1 \\
2454601.810927 & 0.99775 & 0.00134 & 1 \\
2454601.811321 & 0.99772 & 0.00134 & 1 \\
2454601.811703 & 0.99943 & 0.00134 & 1 \\
\enddata

\tablecomments{(1) Fred L. Whipple Observatory 1.2m telescope, (2) University of Hawaii 2.2m telescope. \\ (This table is available in its entirety online.)}

\end{deluxetable}

\tabletypesize{\scriptsize}

\begin{deluxetable}{ccc}
\tablewidth{0pt}
\tablecaption{Keck/HIRES Doppler Shift Measurements of WASP-3}
\tablehead{\colhead{Heliocentric Julian Date} & \colhead{Radial Velocity } & 
\colhead{Measurement Uncertainty } \\ 
\colhead{} & \colhead{(m s$^{-1}$)} & \colhead{(m s$^{-1}$)} } 
\startdata
2454636.833671 & 162.68512 & 6.42080 \\
2454636.878764 & 101.27652 & 6.92383 \\
2454636.899783 & 83.69744 & 7.16853 \\
2454636.908996 & 156.81285 & 7.13308 \\
2454636.918452 & 177.62252 & 6.10846 \\
2454636.930559 & 138.14356 & 7.13066 \\
2454636.940096 & 117.51487 & 6.84713 \\
2454636.951218 & 51.83493 & 6.34730 \\
2454636.959239 & 3.28570 & 6.49503 \\
2454636.969887 & -39.46792 & 7.40002 \\
2454636.980942 & -80.49891 & 6.88382 \\
2454636.991034 & -63.31213 & 7.93881 \\
2454637.001983 & 56.56807 & 8.33331 \\
2454637.014090 & -6.59426 & 6.29524 \\
2454637.060144 & -92.67312 & 7.94984 \\
2454637.090295 & -100.37790 & 6.66956 \\
2454638.816277 & -78.81247 & 7.72656 \\
2454638.825606 & -126.87940 & 7.02801 \\
2454638.836659 & -125.27818 & 7.66176 \\
2454638.845548 & -130.38241 & 6.97449 \\
2454638.854761 & -55.94224 & 7.53164 \\
2454638.865225 & -51.31588 & 7.19155 \\
2454638.877991 & -60.45514 & 7.23436 \\
2454638.890896 & -43.96450 & 6.91079 \\
2454639.006175 & -155.20399 & 6.97899 \\
2454674.740671 & -69.57869 & 7.92163 \\
2454674.744085 & -41.62488 & 9.83204 \\
2454674.868667 & 66.23444 & 8.56491 \\
2454674.994730 & 182.86504 & 9.53239 \\
2454983.909935 & 259.36496 & 9.73989 \\
2454984.890077 & -124.05018 & 8.87455 \\
2454986.041289 & -141.69685 & 8.96584 \\
2454986.920247 & 34.36423 & 8.97649 
\enddata
\label{tab:keckdata}
\end{deluxetable}


\tabletypesize{\normalsize}

\begin{deluxetable}{ccc}

\tablewidth{0pt}
\tablecaption{WASP-3 Midtransit Times \label{tab:tc}}
\tablehead{\colhead{Epoch} & \colhead{Midtransit time [HJD]} & \colhead{Reference}  } 
\startdata
$-250$ & 2,454,143.8503 $\pm$ 0.0004 & 1 \\
$-2$   & 2,454,601.86514 $\pm$ 0.00027 &  2 \\
0  & 2,454,605.55956 $\pm$ 0.00035 & 3 \\
12 & 2,454,627.72098 $\pm$  0.00031 &  2 \\
18 & 2,454,638.80329 $\pm$ 0.00031 &  2 \\
30 & 2,454,660.96435 $\pm$ 0.00021 &  2 \\
59 & 2,454,714.52210 $\pm$ 0.00036 &  3 \\
194 & 2,454,963.84361 $\pm$  0.00072 &  2 \\
201 & 2,454,976.77290 $\pm$ 0.00051 &  2
\enddata

\tablecomments{References: (1) Pollacco et al.~(2008), (2) This work,
  (3) Gibson et al.~(2008).}

\end{deluxetable}


\begin{deluxetable}{cc}

\tablewidth{0pt}
\tablecaption{WASP-3 Photometric Model Parameters\label{tab:photvalues}}
\tablehead{\colhead{Parameter} & \colhead{Photometric value}} 
\startdata
Orbital Period [d]       & $1.846834 \pm 0.000002$  \\
Midtransit time [HJD]    & $2454605.55922 \pm 0.00019 $ \\
Impact parameter, $b$    & $0.531_{-0.043}^{+0.036}$ \\
Approximate transit duration, $T$  [hr] & $2.412_{-0.015}^{+0.017}$ \\
Transit duration [hr] & $2.813\pm 0.012$ \\
$R_p/R_\star$ (UT 2008~May~15, $i'$)  &  $0.1038_{-0.0011}^{+0.0007}$\\
$R_p/R_\star$ (UT 2008~June~9, $i'$) & $0.1041_{-0.0012}^{+0.0008}$\\
$R_p/R_\star$ (UT 2008~June~21, $i'$) & $0.1011_{-0.0011}^{+0.0009}$\\
$R_p/R_\star$ (UT 2008~July~13, $z'$) & $0.1099_{-0.0010}^{+0.0006}$\\
$R_p/R_\star$ (UT 2009~May~12, $g'$) & $0.1107_{-0.0020}^{+0.0014}$\\
$R_p/R_\star$ (UT 2009~May~25, $g'$) & $0.1058_{-0.0016}^{+0.0014}$\\
Limb-darkening coefficient $u_1$ ($g'$) & $0.596_{-0.033}^{+0.031}$\\
Limb-darkening coefficient $u_2$ ($g'$) & $0.215_{-0.027}^{+0.025}$\\
Limb-darkening coefficient $u_1$ ($i'$) & $0.288_{-0.035}^{+0.023}$\\
Limb-darkening coefficient $u_2$ ($i'$) & $0.321_{-0.028}^{+0.019}$\\
Limb-darkening coefficient $u_1$ ($z'$) & $0.193_{-0.042}^{+0.030}$\\
Limb-darkening coefficient $u_2$ ($z'$) & $0.283_{-0.036}^{+0.022}$\\
\enddata

\tablecomments{The values and uncertainties for $P$ and $T_c$ were
  determined by fitting a straight line to the data given in
  Table~\ref{tab:tc} and multiplying the formal errors in the fit by
  $\sqrt{\chi^2_\nu} = \sqrt{3}$.  The transit duration is dependent
  on the radius ratio ($R_p/R_\star$), and hence the value quoted is
  the mean and standard deviation of the 6 transit durations
  determined (using each of the $R_p/R_\star$ values).  For the other
  parameters, the quoted value represents the mode of the MCMC
  distribution, and the quoted uncertainties are ``1$\sigma$'' errors,
  spanning the range between the 15.85\% and 84.15\% levels of the
  MCMC cumulative distribution.  One final note is that the
  approximate transit duration ($T$) refers to the quantity defined in
  Eqn. \ref{eqn:T}, making it different from the transit duration.}

\end{deluxetable}


\begin{deluxetable}{lc}
\tablewidth{0pt}
\tablecaption{WASP-3 RV Model Parameters \label{tab:rvparams}}
\tablehead{\colhead{Parameter} & \colhead{Value}}

\startdata
$v\sin i_\star$ [km \s]  & $14.1 ^{+1.5} _{-1.3}$ \\
$\lambda$ [deg]         & $3.3 ^{+2.5} _{-4.4}$\\
$K$ [m \s] & $290.5^{+9.8} _{-9.2}$\\
$M_{ \rm p}$ [$M_{\rm J}$] \tablenotemark{a} & $2.04^{+0.07} _{-0.07}$\\
$\gamma$ [m \s]         & $33.5 ^{+6.3} _{-4.5}$\\
$\Delta\gamma$ [m \s] & $14.1 ^{+6.2} _{-6.9}$
\enddata
\tablenotetext{a}{Assuming $M_{\star}=1.24~M_{\odot}$ per
  \citet{pollacco}, and $i=84.1^{\circ}$, as determined from our light
  curve analysis.}
\end{deluxetable}


\end{document}